\documentclass[12pt]{amsart} 
\usepackage{amssymb}
\usepackage{amsmath}

\begin{document}

{\normalsize Theoretical and Mathematical Physics, Vol. 124, No. 1, 
pp. 859 -- 871, 2000. Translated from Teoreticheskaya i Matematicheskaya Fizika,
Vol. 124, No. 1, pp. 3--17, July 2000 (in Russian).
\footnote{In this version a few misprints have been corrected.}

\vskip 2.0cm

\title[On the spectrum of the periodic Dirac operator]{On the spectrum of the 
periodic Dirac operator}
\author{L.I.{\,}Danilov}
\date{}

\address{Physical-Technical Institute, Ural Branch of the Russian Academy of Sciences,
Kirov Street 132, Izhevsk, 426000, Russia}
\email{danilov@otf.pti.udm.ru}


\begin{abstract}
The absolute continuity of the spectrum for the periodic Dirac operator
$$
\widehat D=\sum\limits_{j=1}^n\biggl( -i\, \frac {\partial}{\partial x_j}-A_j\biggr)
\widehat \alpha _j+\widehat V^{(0)}+\widehat V^{(1)}, \quad x\in {\mathbb R}^n,
\quad n\geq 3,
$$
is proved given that either $A\in C({\mathbb R}^n;{\mathbb R}^n)\cap H^q_{\mathrm 
{loc}}({\mathbb R}^n;{\mathbb R}^n)$, $2q>n-2$, or the Fourier series of the vector
potential $A:{\mathbb R}^n\to {\mathbb R}^n$ is absolutely convergent. Here,
$\widehat V^{(s)}=(\widehat V^{(s)})^*$ are continuous matrix functions and 
$\widehat V^{(s)}\widehat \alpha _j=(-1)^s\widehat \alpha _j\widehat V^{(s)}$ for all 
anticommuting Hermitian matrices $\widehat \alpha _j$, $\widehat \alpha _j^2=
\widehat I$, $s=0,1$.
\end{abstract}

\maketitle

\large

In [1], the absolute continuity of the spectrum for the periodic Dirac operator
$$
\widehat D=\sum\limits_{j=1}^n\biggl( -i\, \frac {\partial}{\partial x_j}-A_j\biggr)
\widehat \alpha _j+V\widehat I+V_0\widehat \alpha _{n+1}
$$
in ${\mathbb R}^n$, $n\geq 2$, was proved, where $V,V_0\in L^q_{\mathrm {loc}}
({\mathbb R}^2;{\mathbb R})$, $A\in L^q_{\mathrm {loc}}({\mathbb R}^2;{\mathbb R}^2)$,
$q>2$, for $n=2$ and $V,V_0\in C({\mathbb R}^n;{\mathbb R})$, $A\in C^{2n+3}({\mathbb 
R}^n;{\mathbb R}^n)$ for $n\geq 3$. Here, $\widehat \alpha _{n+1}$ is a Hermitian
matrix anticommuting with the matrices $\widehat \alpha _j$, $j=1,\dots ,n$, and
$\widehat \alpha _{n+1}^2=\widehat I$. For $n=2$, the proof is based on the results
in [2,3], where the two-dimensional periodic Schr\"odinger operator
was considered. In [3], the absolute continuity of the spectrum for this operator was
proved in the case of the scalar (electric) and the vector (magnetic) potentials
$V$ and $A$ satisfying the conditions $V\in L^q_{\mathrm {loc}}({\mathbb R}^2;
{\mathbb R})$ and $A\in L^{2q}_{\mathrm {loc}}({\mathbb R}^2;{\mathbb R}^2)$, $q>1$.
For the periodic Dirac operator with $n=2$, the same result as in [1] was
independently obtained in [4]. However, it was assumed in [4] that $V_0\equiv m=
{\mathrm {const}}$. But the functions $V_0\in L^q_{\mathrm {loc}}({\mathbb R}^2;
{\mathbb R})$, $q>2$, can in fact be considered in this case as well without any
significant changes. The proof in [4] used the method suggested in [5], where the
absolute continuity of the spectrum was established for the two-dimensional Dirac
operator with the periodic potential $V\in L^q_{\mathrm {loc}}({\mathbb R}^2;
{\mathbb R})$, $q>2$ (and $A\equiv 0$). Sobolev's results (see [6]) for the absolute 
continuity of the spectrum of the Schr\"odinger operator with the periodic vector
potential $A\in C^{2n+3}({\mathbb R}^n;{\mathbb R}^n)$ were used in [1] for the case
$n\geq 3$. Sobolev later replaced the last condition with the weaker condition $A\in
H^q_{\mathrm {loc}}({\mathbb R}^n;{\mathbb R}^n)$, $2q>3n-2$, $n\geq 3$ (see the
survey in [7]), which permitted changing the smoothness conditions on the vector
potential $A$ for the periodic Dirac operator [1,7] in an adequate manner. The
absolute continuity of the spectrum for the Dirac operator in ${\mathbb R}^n$, $n\geq
3$, with the periodic scalar potential $V$ (for $A\equiv 0$) was proved in [8--10]
under various constraints on $V$.
\vskip 0.2cm

{\bf 1.} Let ${\mathcal L}_M$, $M\in {\mathbb N}$, denote the linear space of complex
$M\times M$ matrices, let ${\mathcal S}_M$ be the set of Hermitian matrices in
${\mathcal L}_M$, and let the matrices $\widehat \alpha _j\in {\mathcal S}_M$,
$j=1,\dots ,n$, satisfy the commutation relations $\widehat \alpha _j\widehat \alpha 
_l+\widehat \alpha _l\widehat \alpha _j=2\delta _{jl}\widehat I$, where $\widehat
I\in {\mathcal L}_M$ is the identity matrix and $\delta _{jl}$ is the Kronecker
delta. We write
$$
{\mathcal L}_M^{(s)}=\{ \widehat L\in {\mathcal L}_M:\widehat L\widehat \alpha _j=
(-1)^s\widehat \alpha _j\widehat L\ \, \text{for\ all} \ \, j=1,\dots ,n\} ,
$$ $$
{\mathcal S}_M^{(s)}={\mathcal L}_M^{(s)}\cap {\mathcal S}_M, \quad s=0,1.
$$

We consider the Dirac operator
$$
\widehat D=\widehat D_0+\widehat V^{(0)}+\widehat V^{(1)}-\sum\limits_{j=1}^n
A_j\widehat \alpha _j=\sum\limits_{j=1}^n\biggl( -i\, \frac {\partial}{\partial x_j}
-A_j\biggr) \widehat \alpha _j+\widehat V^{(0)}+\widehat V^{(1)},  \eqno (1)
$$
where $n\geq 3$ ($i^2=-1$). The vector function $A:{\mathbb R}^n\to {\mathbb R}^n$
and the matrix functions $\widehat V^{(s)}:{\mathbb R}^n\to {\mathcal S}_M^{(s)}$,
$s=0,1$, are assumed to be periodic with a period lattice $\Lambda \subset
{\mathbb R}^n$. We set
$$
\widehat V=\widehat V^{(0)}+\widehat V^{(1)}-\sum\limits_{j=1}^nA_j\widehat \alpha 
_j.
$$
The coordinates of the vectors in ${\mathbb R}^n$ are set in an orthogonal basis
$\{ {\mathcal E}_j\} $. Here, $E_j$ and $E_j^*$ are the basis vectors in the
lattice $\Lambda $ and its reciprocal lattice $\Lambda ^*$, $(E_j,E_l^*)=\delta _{jl}$
($|.|$ and $(.,.)$ are the length and the inner product of vectors in ${\mathbb 
R}^n$),
$$
K=\biggl\{ x=\sum\limits_{j=1}^n\xi _jE_j:0\leq \xi _j<1, j=1,\dots ,n\biggr\} ,
$$ $$
K^*=\biggl\{ y=\sum\limits_{j=1}^n\eta _jE_j:0\leq \eta _j<1, j=1,\dots ,n\biggr\} ,
$$
and $v(K)$ and $v(K^*)$ are the volumes of the elementary cells $K$ and $K^*$.

The inner products and the norms in the spaces $L^2(K;{\mathbb C}^M)$ and ${\mathbb 
C}^M$ are introduced in the usual way with (as a rule) the usual notation (without
indicating the spaces themselves). The matrices in ${\mathcal L}_M$ are identified
with the operators on the space ${\mathbb C}^M$ (and their norm is defined as the norm
of operators on ${\mathbb C}^M$). Let $H^q({\mathbb R}^n;{\mathbb C}^d)$, $d\in
{\mathbb N}$, be the Sobolev class of order $q\geq 0$, and let $\widetilde H^q
(K;{\mathbb C}^d)$ be the set of vector functions $\phi :K\to {\mathbb C}^d$ whose
periodic extensions (with the period lattice $\Lambda $) belong to $H^q_{\mathrm
{loc}}({\mathbb R}^n;{\mathbb C}^d)$. In what follows, the functions defined on the
elementary cell $K$ are identified with their periodic extensions throughout the
space ${\mathbb R}^n$.

We let
$$
\chi _N=v^{-1}(K)\int\limits_K\chi (x)\, e^{-2\pi i\, (N,x)}\, d^nx,\quad
N\in \Lambda ^*,
$$
denote the Fourier coefficients of the functions $\chi \in L^1(K,U)$, where $U$ is
the space ${\mathbb C}$ or ${\mathbb C}^M$ or ${\mathcal L}_M$.

Let ${\mathcal B}({\mathbb R})$ be the set of Borel subsets ${\mathcal O}\subseteq
{\mathbb R}$, and let ${\mathcal M}_h$, $h>0$, be the set of signed even Borel 
measures (charges) $\mu :{\mathcal B}({\mathbb R})\to {\mathbb R}$ such that
$$
\widehat \mu (p)=\int\limits_{{\mathbb R}}e^{\, ipt}\, d\mu (t)=1\ \, \text{for}
\ \, |p|\leq 2\pi h,\quad p\in {\mathbb R},
$$ $$
\| \mu \| =\sup\limits_{{\mathcal O}\, \in \, {\mathcal B}({\mathbb R})}\bigl(
|\mu ({\mathcal O})|+|\mu ({\mathbb R}\backslash {\mathcal O})|\bigr) <+\infty ,
\quad \mu \in {\mathcal M}_h.
$$
For an arbitrary vector $\gamma \in \Lambda \backslash \{ 0\} $, an arbitrary measure
$\mu \in {\mathcal M}_h$, $h>0$, and any vector $\widetilde e\in S_{n-2}(|\gamma |
^{-1}\gamma )=\{ e^{\prime }\in S_{n-1}:(\gamma ,e^{\prime })=0\} $, where $S_{n-1}$
is the unit sphere in ${\mathbb R}^n$, we write
$$
\widetilde A(\gamma ,\mu ,\widetilde e;x)=\int\limits_{{\mathbb R}}d\mu (t)
\int\limits_0^1A(x-\xi \gamma -t\widetilde e\, )\, d\xi ,\quad x\in {\mathbb R}^n.
$$

In this paper, we consider continuous (periodic) functions $A:{\mathbb R}^n\to
{\mathbb R}^n$ and $\widehat V^{(s)}:{\mathbb R}^n\to {\mathcal S}_M^{(s)}$, $s=0,1$.
In this case, $\widehat D=\widehat D_0+\widehat V$ is a self-adjoint operator on the
Hilbert space $L^2({\mathbb R}^n;{\mathbb C}^M)$ with the domain $D(\widehat D)=
D(\widehat D_0)=H^1({\mathbb R}^n;{\mathbb C}^M)$.
\vskip 0.2cm

{\bf Theorem 1.} {\it Let $A:{\mathbb R}^n\to {\mathbb R}^n$ and $\widehat V^{(s)}:
{\mathbb R}^n\to {\mathcal S}_M^{(s)}$, $s=0,1$, be continuous periodic functions
with the period lattice $\Lambda \subset {\mathbb R}^n$, $n\geq 3$. If
$$
\max\limits_{\widetilde e\, \in \, S_{n-2}(|\gamma |^{-1}\gamma )}\, \bigl\| 
|\widetilde A(\gamma ,\mu ,\widetilde e;.)-A_0| \bigr\| _{L^{\infty }({\mathbb 
R}^n)}<\pi |\gamma |^{-1}  \eqno (2)
$$
for some vector $\gamma \in \Lambda \backslash \{ 0\}$ and a measure $\mu \in
{\mathcal M}_h$, $h>0$, where
$$
A_0=v^{-1}(K)\int\limits_KA(x)\, d^nx,
$$
then the spectrum of operator $\mathrm (1)$ is absolutely continuous.}
\vskip 0.2cm

The operator $\widehat D$ is unitarily equivalent to the direct integral
$$
{\int\limits_{2\pi K^*}}^{\bigoplus }\ \widehat D(k)\, \frac {d^nk}{(2\pi )^nv(K^*)}
\, ,  \eqno (3)
$$
where
$$
\widehat D(k)=\widehat D_0(k)+\widehat V, \quad \widehat D_0(k)=\sum\limits_{j=1}^n
\biggl( -i\, \frac {\partial }{\partial x_j}+k_j\biggr) \widehat \alpha _j, \quad
k_j=(k,{\mathcal E}_j),
$$ $$
D(\widehat D(k))=D(\widehat D_0(k))=\widetilde H^1(K;{\mathbb C}^M)\subset L^2(K;
{\mathbb C}^M).
$$
The vector $k\in {\mathbb R}^n$ is called a quasimomentum. The unitary equivalence
is established using the Gel'fand transformation [11] (also see [9] for the case
of the periodic Dirac operator). The self-adjoint operators $\widehat D(k)$ have
compact resolvents and hence discrete spectra. Let $E_{\nu }(k)$, $\nu \in {\mathbb 
Z}$, be the eigenvalues of the operators $\widehat D(k)$. We assume that they are
arranged in an increasing order (counting multiplicities). The eigenvalues can be
indexed for different $k$ such that the functions ${\mathbb R}^n\ni k\to E_{\nu }(k)$
are continuous.

Let $e\in S_{n-1}$. For $k\in {\mathbb R}^n$ and $\varkappa \geq 0$, we write
$$
\widehat D_0(k+i\varkappa e)=\widehat D_0(k)+i\varkappa \sum\limits_{j=1}^ne_j
\widehat \alpha _j\, , \quad e_j=(e,{\mathcal E}_j),
$$ $$
\widehat D(k+i\varkappa e)=\widehat D_0(k+i\varkappa e)+\widehat V,
$$ $$
D(\widehat D(k+i\varkappa e))=D(\widehat D_0(k+i\varkappa e))=\widetilde H^1(K;
{\mathbb C}^M).
$$
\vskip 0.2cm

{\bf Proof of Theorem 1.} We use the Thomas method [12]. Because it is well known
[4,9] (see [2,13] for the case of the periodic Schr\"odinger operator), we present
only a brief scheme of the method. The decomposition of the operator $\widehat D$
into direct integral (3) and the piecewise analyticity of the functions ${\mathbb R}
\ni \xi \to E_{\nu }(k+\xi e)$, $\nu \in {\mathbb Z}$, $k\in {\mathbb R}^n$, imply
(see Theorems XIII.85 and XIII.86 in [13]) that to prove the absolute continuity
of the spectrum of operator (1), it suffices to show that the functions $\xi \to
E_{\nu }(k+\xi e)$ are not constant (for some unit vector $e$) on every interval
$(\xi _1,\xi _2)\subset {\mathbb R}$. But if we suppose that $E_{\nu }(k+\xi e)
\equiv E$ for all $\xi \in (\xi _1,\xi _2)$, $\xi _1<\xi _2$, then it follows from 
the analytic Fredholm theorem that $E$ is an eigevalue of $\widehat D(k+(\xi +i
\varkappa )e)$ for all $\xi +i\varkappa \in {\mathbb C}$. Consequently, it suffices
to prove the invertibility of the operators $\widehat D(k+(\xi +i\varkappa )e)-E$,
$k\in {\mathbb R}^n$, $E\in {\mathbb R}$, for some $\xi +i\varkappa \in {\mathbb C}$.
Theorem 1 is therefore a consequence of the following assertion. 
\vskip 0.2cm

{\bf Theorem 2.} {\it Let $\gamma \in \Lambda \backslash \{ 0\} $, $e=|\gamma |^{-1}
\gamma $, $\mu \in {\mathcal M}_h$, $h>0$. Let $A:{\mathbb R}^n\to {\mathbb C}^n$
and $\widehat V^{(s)}:{\mathbb R}^n\to {\mathcal L}_M^{(s)}$, $s=0,1$, be
continuous periodic functions with the period lattice $\Lambda \subset {\mathbb R}^n$,
$n\geq 3$. If $A_0=0$ and
$$
\max\limits_{\widetilde e\, \in \, S_{n-2}(|\gamma |^{-1}\gamma )}\, \bigl\| (
\widetilde A(\gamma ,\mu ,\widetilde e;.),\widetilde e\, )+i(\widetilde A(\gamma ,\mu 
,\widetilde e;.),e)\bigr\| _{L^{\infty }({\mathbb R}^n)}=\widetilde \theta \pi
|\gamma |^{-1},
$$
where $\widetilde \theta \in [0,1)$, then for any $\theta \in (0,1-\widetilde \theta 
\, )$, there exists a number $\varkappa _0=\varkappa _0(\gamma ,h,\mu ;\widehat V,
\theta )>0$ such that the inequality
$$
\| \widehat D(k+i\varkappa e)\phi \| \geq \theta \pi |\gamma |^{-1} \exp \, (-4C\, \| 
\mu \| \, \max \{ |\gamma |,h^{-1}\} \, \| A\| _{L^{\infty }({\mathbb R}^n;{\mathbb 
C}^n)})\, \| \phi \|
$$
holds for all $k\in {\mathbb R}^n$ with $(k,\gamma )=\pi $, all $\varkappa \geq
\varkappa _0$, and all vector functions $\phi \in \widetilde H^1(K;{\mathbb C}^M)$,
where $C>0$ is a universal constant to be defined in Lemma 1.}
\vskip 0.2cm

Theorem 2 is proved in Section 3. The following theorem is a consequence of Theorem 1.
\vskip 0.2cm

{\bf Theorem 3.} {\it Let $A:{\mathbb R}^n\to {\mathbb R}^n$ and $\widehat V^{(s)}:
{\mathbb R}^n\to {\mathcal S}_M^{(s)}$, $s=0,1$, be continuous periodic functions 
with the period lattice $\Lambda \subset {\mathbb R}^n$, $n\geq 3$. If at least
one of the conditions

1. $A\in H^q_{{\mathrm {loc}}}({\mathbb R}^n;{\mathbb R}^n)$, $2q>n-2$, or

2. $\sum\limits_{N\, \in \, \Lambda ^*}\, \| A_N\| _{{\mathbb C}^n}<+\infty $ \\
holds, then the spectrum of operator $\mathrm (1)$ is absolutely continuous.}
\vskip 0.2cm

Theorem 4 is used to prove Theorem 3.
\vskip 0.2cm

{\bf Theorem 4.} {\it Let $\Lambda $ be a lattice in ${\mathbb R}^n$, $n\geq 2$.
There are positive constants $c_1$ and $c_2$ depending on $n$ and $\Lambda $ such
that for any nonnegative Borel measure $\mu $ on the unit sphere $S_{n-1}\subset
{\mathbb R}^n$, any $h>0$, and any $R_0\geq \min\limits_{\gamma \, \in \, \Lambda
\backslash \{ 0\} }|\gamma |$, there exists a vector $\gamma \in \Lambda \backslash 
\{ 0\} $ such that

1. $|\gamma |\leq R_0$,

2. if $(\gamma ,\gamma ^{\, \prime })=0$ for some vector $\gamma ^{\, \prime }\in 
\Lambda \backslash \{ 0\} $, then $|\gamma ^{\, \prime }|>c_1R_0^{\, 1/(n-1)}$
$\mathrm ($$\Lambda ^*$ is the reciprocal lattice of $\Lambda $$\mathrm )$,

3. $\mu \bigl( \{ e^{\, \prime }\in S_{n-1}:|(e^{\, \prime },\gamma )|\leq h\} \bigr)
\leq c_2|\gamma |^{-1}\max \, \bigl\{ h,R_0^{-1/(n-1)}\bigr\} \, \mu (S_{n-1})$.}
\vskip 0.2cm

The proof of Theorem 4 for the lattice $\Lambda ={\mathbb Z}^n$ and for $h=c_3
R_0^{-1/(n-1)}$ (where $c_3=c_3(n)>0$) is presented in [14] (see [15] for $n=3$).
The proof in the general case follows the one suggested in [14] with some slight
changes.
\vskip 0.2cm

{\bf Proof of Theorem 3.} It can be assumed that $A_0=0$. We write
$$
F(A;\gamma ,\mu )=\max\limits_{\widetilde e\, \in \, S_{n-2}(|\gamma |^{-1}\gamma )}\,
|\gamma |\, \| \widetilde A(\gamma ,\mu ,\widetilde e;.)\| _{L^{\infty }({\mathbb
R}^n;{\mathbb R}^n)}, \quad \gamma \in \Lambda \backslash \{ 0\} , \ \ \mu \in
{\mathcal M}_h.
$$
Let condition 1 hold. We define the measure
$$
\mu ^{(1)}(.)=\sum\limits_{N\, \in \, \Lambda ^*\backslash \{ 0\} }\, |N|^{2q}\, 
\| A\| ^2_{{\mathbb C}^n}\delta _{N/|N|}(.)
$$
on the unit sphere $S_{n-1}$, where $\delta _{e^{\, \prime }}(.)$ is the Dirac
measure concentrated at the point $e^{\, \prime }\in S_{n-1}$. From Theorem 4
(applied to the measure $\mu ^{(1)}$), it follows that for any $R_0\geq \min\limits
_{\gamma \, \in \, \Lambda \backslash \{ 0\} }|\gamma |$ there is a vector $\gamma
\in \Lambda \backslash \{ 0\} $ such that $|\gamma |\leq R_0$,
$$
\sum\limits_{N\, \in \, \Pi (\gamma )}|N|^{2q}\, \| A_N\| ^2_{{\mathbb C}^n}\leq
c_2|\gamma |^{-1}R_0^{-1/(n-1)}\sum\limits_{N\, \in \, \Lambda ^*}|N|^{2q}\, \| A_N\| 
^2_{{\mathbb C}^n},
$$
and $|\gamma ^{\, \prime }|>c_1R_0^{1/(n-1)}$ for all $\gamma ^{\, \prime }\in \Pi
(\gamma )\doteq \{ \gamma ^{\, \prime }\in \Lambda ^*\backslash \{ 0\} :(\gamma ,
\gamma ^{\, \prime })=0\} $. We take a measure $\mu \in {\mathcal M}_h$ (for some
$h>0$) such that $|\widehat \mu (p)|\leq 1$ for all $p\in {\mathbb R}$ and
$\widehat \mu (p)=0$ if $|p|\geq 2\pi h_1>2\pi h$. For a vector $\widetilde e\in
S_{n-2}(|\gamma |^{-1}\gamma )$, we write $\Pi (\gamma ,\widetilde e\, )=\{ \gamma ^{\,
\prime }\in \Pi (\gamma ):|(\gamma ^{\, \prime },\widetilde e\, )|\leq h_1\} $. 
Because $2q>n-2$, we have
$$
\sum\limits_{N\, \in \, \Pi (\gamma ,\widetilde e\, )}|N|^{-2q}\leq c_4R_0^{-2q/(n-1)}
$$
for all $\widetilde e\in S_{n-2}(|\gamma |^{-1}\gamma )$, where the constant $c_4
>0$ depends on $n$, $\Lambda $, $q$, and $h_1$. Consequently,
$$
F(A;\gamma ,\mu )\leq \sup\limits_{\widetilde e\, \in \, S_{n-2}(|\gamma |^{-1}
\gamma )}\, |\gamma |\, \sum\limits_{N\, \in \, \Pi (\gamma ,\widetilde e\, )}\, \| 
A_N\| _{{\mathbb C}^n}\leq  \eqno (4)
$$ $$
|\gamma |\, \biggl( \, \sup\limits_{\widetilde e\, \in \, S_{n-2}(|\gamma |^{-1}
\gamma )}\ \sum\limits_{N\, \in \, \Pi (\gamma ,\widetilde e\, )}|N|^{-2q}\biggr) 
^{1/2}\biggl( \, \sum\limits_{N\, \in \, \Pi (\gamma )}\, |N|^{2q}\, \| A_N\| 
_{{\mathbb C}^n}^2\biggr) ^{1/2} \leq
$$ $$ 
\sqrt {c_2c_4}\, R_0^{\, (n-2-2q)/(2(n-1))}\, \biggl( \, \sum\limits_{N\, \in \, 
\Lambda ^*}\, |N|^{2q}\, \| A_N\| _{{\mathbb C}^n}^2\biggr) ^{1/2}.
$$
The right-hand side of (4) becomes arbitrarily small if a sufficiently large number
$R_0$ is chosen (and inequality (2) consequently holds). Case 2, for which the Dirac
measure $\mu =\delta $ is chosen, is considered in a similar (slightly simpler) way.
Theorem 3 is proved.
\vskip 0.2cm

{\bf 2.} We fix a vector $\gamma \in \Lambda \backslash \{ 0\} $ and a measure
$\mu \in {\mathcal M}_h$, $h>0$, $e=|\gamma |^{-1}\gamma $. In what follows, the
constants we introduce can depend on $\gamma $, $h$, and $\mu $, but we do not
indicate this dependence explicitly (untill Theorem 8 below).

Let $\widehat P^{\, {\mathcal C}}$, where ${\mathcal C}\subseteq \Lambda ^*$, denote
the orthogonal projection on $L^2(K;{\mathbb C}^M)$ that takes a vector function
$\phi \in L^2(K;{\mathbb C}^M)$ to the vector function
$$
\widehat P^{\, {\mathcal C}}\phi =\phi ^{{\mathcal C}}=\sum\limits_{N\, \in \,
{\mathcal C}}\phi _N\, e^{\, 2\pi i\, (N,x)}
$$
(here, $\phi ^{\, \emptyset }\equiv 0$). We introduce the notation ${\mathcal H}
({\mathcal C})=\{ \phi \in L^2(K;{\mathbb C}^M):\phi _N=0\, $ for $N\notin {\mathcal 
C}\} $.

Let ${\mathcal P}(e)=\{ \tau e:\tau \in {\mathbb R}\} $. For the vectors $x\in
{\mathbb R}^n\backslash {\mathcal P}(e)$, we write
$$
\widetilde e(x)=(x-(x,e)e)\, |x-(x,e)e|^{-1}\in S_{n-2}(e),
$$
where $S_{n-2}(e)=\{ \widetilde e\in S_{n-1}:(e,\widetilde e\, )=0\} $; we also write
$\sigma _{n-2}={\mathrm {mes}}\, (S_{n-2})$, where ${\mathrm {mes}}\, (.)$ is the
standard measure (`surface area') on the unit sphere $S_{n-2}=S_{n-2}(e)$. For
$\beta >0$ and $\varkappa >\beta $, we write
$$
{\mathcal O}_{\beta }={\mathcal O}_{\beta }(\varkappa )=\{ x\in {\mathbb R}^n:
|(x,e)|<\beta \ \, \text{and}\, \ \bigl| \varkappa -|x-(x,e)e|\bigr| <\beta \} ,
$$ $$
{\mathcal K}_{\beta }={\mathcal K}_{\beta }(k;\varkappa )=\{ N\in \Lambda ^*:
k+2\pi N\in {\mathcal O}_{\beta } \}, \quad k\in {\mathbb R}^n.
$$
We set
$$
\widehat P^{\, \pm }_{\widetilde e}=\frac 12\, \biggl( \widehat I\mp i\, \biggl( \,
\sum\limits_{j\, =\, 1}^ne_j\widehat \alpha _j\biggr) \biggl( \, \sum\limits_{j\, =\, 
1}^n\widetilde e_j\widehat \alpha _j\biggr) \biggr)
$$
for all $\widetilde e\in S_{n-2}(e)$, where $\widehat P^{\, \pm }_{\widetilde e}$
are orthogonal projections on ${\mathbb C}^M$.

For $k\in {\mathbb R}^n$, $\varkappa \geq 0$, and $N\in \Lambda ^*$, we introduce
the notation
$$
\widehat D_N(k;\varkappa )=\sum\limits_{j\, =\, 1}^n(k_j+2\pi N_j+i\varkappa e_j)\, 
\widehat \alpha _j\, ,
$$ $$
G^{\, \pm }_N(k;\varkappa )=\biggl( (k+2\pi N,e)^2+\bigl( \varkappa \pm \sqrt
{|k+2\pi N|^2-(k+2\pi N,e)^2}\, \bigr) ^2 \biggr) ^{1/2},
$$
and $G_N(k;\varkappa )=G^{\, -}_N(k;\varkappa )$. The inequalities 
$$
G_N(k;\varkappa )\, \| u\| \leq \| \widehat D_N(k;\varkappa )u\| \leq G^{\, +}_N
(k;\varkappa )\, \| u\| ,\quad u\in {\mathbb C}^M,
$$
hold. If $(k,\gamma )=\pi $, then $G_N(k;\varkappa )\geq |(k+2\pi N,e)|\geq \pi
|\gamma |^{-1}$. For all vector functions $\phi \in \widetilde H^1(K;{\mathbb C}^M)$,
$$
\widehat D_0(k+i\varkappa e)\phi =\sum\limits_{N\, \in \, \Lambda ^*}\widehat D_N
(k;\varkappa )\, \phi _N\, e^{\, 2\pi i\, (N,x)}.
$$
In this case (for all $\varkappa \geq 0$ and $k+2\pi N\notin {\mathcal P}(e)$), we
have
$$
\| \widehat D_N(k;\varkappa )\, \widehat P^{\, \pm }_{\widetilde e(k+2\pi N)}\, 
\phi _N\| =G^{\, \pm }_N(k;\varkappa )\, \| \widehat P^{\, \pm }_{\widetilde e(k+2\pi 
N)}\phi _N\| \, ,
$$
and
$$
\widehat P^{\, \pm }_{\widetilde e(k+2\pi N)}\, \widehat D_N(k;\varkappa )\, 
\widehat P^{\, \pm }_{\widetilde e(k+2\pi N)}=\widehat O,
$$
where $\widehat O\in {\mathcal L}_M$ is the zero matrix.

We let $\widehat P^{\, \pm }=\widehat P^{\, \pm }(k)$, where $k\in {\mathbb R}^n$,
denote the operators on $L^2(K;{\mathbb C}^M)$ that take vector functions $\phi \in
L^2(K;{\mathbb C}^M)$ to the vector functions $\widehat P^{\, \pm }\phi \in L^2(K;
{\mathbb C}^M)$ with the Fourier coefficients $(\widehat P^{\, \pm }\phi )_N=
\widehat P^{\, \pm }_{\widetilde e(k+2\pi N)}\phi _N$ if $k+2\pi N\notin {\mathcal 
P}(e)$ and $(\widehat P^{\, \pm }\phi )_N=0$ otherwise.

For the matrix function $\widehat V=\widehat V^{(0)}+\widehat V^{(1)}-\sum\limits_{j=1}
^nA_j\widehat \alpha _j$, where $\widehat V^{(s)}:{\mathbb R}^n\to {\mathcal L}^{(s)}
_M$, $s=0,1$, and $A:{\mathbb R}^n\to {\mathbb C}^n$ are continuous periodic
functions with the period lattice $\Lambda $, we write
$$
W=W(\widehat V)=n\, \| A\| _{L^{\infty }({\mathbb R}^n;{\mathbb C}^n)}+\sum\limits
_{s\, =\, 0,1}\| \widehat V^{(s)}\| _{L^{\infty }({\mathbb R}^n;{\mathcal L}_M)}\, .
$$
We set $c_5(A)=c_5(A;\gamma ,h,\mu )=\exp \, \bigl( -4C\, \| \mu \| \, \max \, \{ 
|\gamma |,h^{-1}\} \, \| A\| _{L^{\infty }({\mathbb R}^n;{\mathbb C}^n)}\bigr) $,
where $C>0$ is a universal constant to be defined in Lemma 1.
\vskip 0.2cm

{\bf Theorem 5.} {\it Let $\widetilde \theta \in [0,1)$, $\theta \in (0,1-\widetilde 
\theta )$, $W_0\geq 0$, $R\geq 1$, $\beta >0$, and $a\in (0,1]$. Also, let us fix a 
vector $\gamma \in \Lambda \backslash \{ 0\} $, a number $h>0$, and a measure $\mu \in
{\mathcal M}_h$$\mathrm ;$ $e=|\gamma |^{-1}\gamma $. Then there are numbers $b=b\, 
(\widetilde \theta , \theta ,W_0;a)>0$ and $\varkappa _0=\varkappa _0(\widetilde 
\theta , \theta ,W_0,R,\beta ;a)>4\beta +R$ such that the inequality
$$
\| \widehat P^{\, +}(k)\, \widehat D(k+i\varkappa e)\, \phi \| ^2+a^2\, \| \widehat 
P^{\, -}(k)\, \widehat D(k+i\varkappa e)\, \phi \| ^2 \geq
$$ $$
c_5^2(A)\, \biggl( \biggl( \theta \, \frac {\pi }{|\gamma |}\biggr) ^2\| P^{\, -}(k)\, 
\phi \| ^2+\biggl( \frac {b\varkappa }{\beta +R}\biggr) ^2\| P^{\, +}(k)\, \phi \| ^2 
\biggr)
$$
holds for all vectors $k\in {\mathbb R}^n$ with $(k,\gamma )=\pi $, all $\varkappa
\geq \varkappa _0$, all continuous periodic functions $\widehat V^{(s)}:{\mathbb R}^n
\to {\mathcal L}^{(s)}_M$, $s=0,1$, and $A:{\mathbb R}^n\to {\mathbb C}^n$ $\mathrm 
($with the period lattice $\Lambda \subset {\mathbb R}^n$, $n\geq 3$$\mathrm )$ such
that $A_0=0$,
$$
W(\widehat V)\leq W_0\, ,  \eqno (5)
$$ $$
\max\limits_{\widetilde e\, \in \, S_{n-2}(e)}\, \| (\widetilde A(\gamma ,\mu ,
\widetilde e;.),\widetilde e\, )+i(\widetilde A(\gamma ,\mu ,\widetilde e;.),e)\|
_{L^{\infty }({\mathbb R}^n)}\leq \widetilde \theta \, \pi |\gamma |^{-1},  \eqno (6)
$$ $$
\widehat V_N=0\ \, \text{for}\, \ 2\pi |N|>R\, ,  \eqno (7)
$$
and all vector functions $\phi \in {\mathcal H}({\mathcal K}_{\beta }(k;\varkappa 
))$.}
\vskip 0.2cm

{\bf Proof.} Without loss of generality we assume that the basis vector ${\mathcal
E}_2$ coincides with $e$. We fix some numbers $\theta <\theta _4<\theta _3<\theta _2
<\theta _1<1-\widetilde \theta $ and write $\delta =1-\theta _4^2\, \theta _3^{-2}$
and $c_5^{\, \prime }=\exp \, \bigl( -4C\, \| \mu \| \, \max \, \{ |\gamma |,h^{-1}\} 
\, W_0\bigr) $. We choose a number $\widetilde \varepsilon \in (0,1)$ proceeding
from the condition $(c_5^{\, \prime })^2\bigl( (1-\widetilde \varepsilon )\, \theta _4
^2-\theta ^2\bigr) \pi ^2|\gamma |^{-2}\geq 2\delta ^{-1}W_0^2\, \widetilde \varepsilon
$. Lower bounds for the constant $\varkappa _0$ are specified in the course of the
proof. We first suppose that $\varkappa _0>4\beta +R$. In this case, if $N\in
{\mathcal K}_{\beta }(k;\varkappa )$, $k\in {\mathbb R}^n$, $\varkappa \geq \varkappa
_0$, and $2\pi |N^{\prime }|\leq R$ (where $N^{\prime }\in \Lambda ^*$), then
$|\widetilde e(k+2\pi (N+N^{\prime }))-\widetilde e(k+2\pi N)|<2R/\varkappa $. There
is a number $c_6=c_6(\widetilde \varepsilon )>0$ such that for all $\varkappa \geq 
\varkappa _0$, there are nonintersecting (nonempty) open sets $\widetilde \Omega
_{\lambda }=\widetilde \Omega _{\lambda }(\varkappa )\subset S_{n-2}=S_{n-2}(e)$ and
vectors $E^{\lambda }=E^{\lambda }(\varkappa )\in \widetilde \Omega _{\lambda }\, $,
$\lambda =1,\dots ,\lambda (\widetilde \varepsilon ,R;\varkappa )$, such that

1. $\ |\widetilde e-E^{\lambda }|\leq \widetilde \rho = c_6R/\varkappa $ for all
$\widetilde e\in \widetilde \Omega _{\lambda }\, $;

2. $\ |\widetilde e^{\, \prime }-\widetilde e^{\, \prime \prime }|> 8R/\varkappa $
for all $\widetilde e^{\, \prime }\in \widetilde \Omega _{\lambda _1}\, $,
$\widetilde e^{\, \prime \prime }\in \widetilde \Omega _{\lambda _2}\, $, $\lambda
_1\neq \lambda _2\, $;

3. $\ {\mathrm {mes}}\, \bigl( S_{n-2}\backslash \bigcup\limits_{\lambda }\widetilde 
\Omega _{\lambda }\bigr) <(1/2)\, \widetilde \varepsilon \, \sigma _{n-2}\, $.

We introduce the notation $\rho =\widetilde \rho +2R/\varkappa $, $\rho ^{\, \prime }=
\widetilde \rho +4R/\varkappa $. Let
$$
\Omega _{\lambda }=\biggl\{ \widetilde e\in S_{n-2}:|\widetilde e-\widetilde e^{\, 
\prime }|< \frac {2R}{\varkappa }\ \, \text{for\ some}\, \ \widetilde e^{\, \prime }
\in \widetilde \Omega _{\lambda }\biggr\} \, ;
$$
$\widetilde \Omega _{\lambda }\subset \Omega _{\lambda }$, and $|\widetilde e^{\, 
\prime }-\widetilde e^{\, \prime \prime }|> 4R/\varkappa $ for all $\widetilde e^{\, 
\prime }\in \Omega _{\lambda _1}\, $, $\widetilde e^{\, \prime \prime }\in 
\Omega _{\lambda _2}\, $, $\lambda _1\neq \lambda _2\, $. Property 3
implies that for any $k\in {\mathbb R}^n$ with $(k,\gamma )=\pi $, any $\varkappa
\geq \varkappa _0\, $, and any $\phi \in {\mathcal H}({\mathcal K}_{\beta }(k;\varkappa 
))$, there is an orthogonal transformation $\widehat S=\widehat S(k,\varkappa ;\phi )$
of the unit sphere $S_{n-2}$ such that (for each of the signs)
$$
\sum\limits_{N\, \in \, {\mathcal K}_{\beta }\, :\, \widetilde e(k+2\pi N)\, \notin
\, \bigcup\limits_{\lambda }\widehat S\widetilde \Omega _{\lambda }}\| \widehat P
^{\, \pm }_{\widetilde e(k+2\pi N)}\phi _N\| ^2\leq \widetilde \varepsilon \, v^{-1}
(K)\, \| \widehat P^{\, \pm }\phi \| ^2.
$$
We write
$$
\widetilde e^{\lambda }=\widehat S(k,\varkappa ;\phi )E^{\lambda },
$$ $$
\widetilde {\mathcal K}^{\lambda }_{\beta }=\widetilde {\mathcal K}^{\lambda }
_{\beta }(k,\varkappa ;\phi )=\{ N\in {\mathcal K}_{\beta }(k;\varkappa ):
\widetilde e(k+2\pi N)\in \widehat S\widetilde \Omega _{\lambda }\} ,
$$ $$
{\mathcal K}^{\lambda }_{\beta }={\mathcal K}^{\lambda }_{\beta }(k,\varkappa ;\phi )
=\{ N\in {\mathcal K}_{\beta }(k;\varkappa ):\widetilde e(k+2\pi N)\in \widehat S
\Omega _{\lambda }\} ,\qquad \widetilde {\mathcal K}^{\lambda }_{\beta }\subset
{\mathcal K}^{\lambda }_{\beta }\, .
$$
The choice of the orthogonal transformation $\widehat S$ means that
$$
\bigl\| (\widehat P^{\, \pm }\phi )^{\, {\mathcal K}_{\beta }\backslash 
\bigcup\limits_{\lambda }\widetilde {\mathcal K}^{\lambda }_{\beta }}\, \bigr\| ^2
\leq \widetilde \varepsilon \, \| \widehat P^{\, \pm }\phi \| ^2.  \eqno (8)
$$ 
For each index $\lambda $ (and for all already chosen $k$, $\varkappa $, and $\phi $),
we take an orthogonal system of vectors ${\mathcal E}_j^{(\lambda )}\in S_{n-1}\, $,
$j=1,\dots ,n$, such that ${\mathcal E}_1^{(\lambda )}=\widetilde e^{\lambda }$ and
${\mathcal E}_2^{(\lambda )}={\mathcal E}_2=e$. We let $x_j^{(\lambda )}=(x,
{\mathcal E}_j^{(\lambda )})$ denote the coordinates of the vectors $x=\sum\limits
_{j=1}^nx_j{\mathcal E}_j\in {\mathbb R}^n$ (and also of the vectors in ${\mathbb
C}^n$). Let ${\mathcal E}_j^{(\lambda )}=\sum\limits_{l=1}^nT_{lj}^{(\lambda )}
{\mathcal E}_l\, $. Then $A_j^{(\lambda )}=\sum\limits_{l=1}^nT_{lj}^{\, (\lambda )}
A_l$ (where $A_l=(A,{\mathcal E}_l)$ and $A_j^{(\lambda )}=(A,{\mathcal E}_j
^{(\lambda )})$), $\widetilde A_j^{(\lambda )}=\widetilde A_j^{(\lambda )}(\gamma ,
\mu ,\widetilde e^{\lambda };.)=\sum\limits_{l=1}^nT_{lj}^{\, (\lambda )}\widetilde
A_l\, $, and $\widetilde A_l=\widetilde A_l(\gamma ,\mu ,\widetilde e^{\lambda };.)$.
We introduce the notation $\widehat \alpha _j^{(\lambda )}=\sum\limits_{l=1}^nT_{lj}
^{\, (\lambda )}\widehat \alpha _l\, $, $j=1,\dots ,n$. For the Fourier coefficients
$(\widetilde A_j^{(\lambda )})_N$ of the functions $\widetilde A_j^{(\lambda )}$,
$j=1,\dots ,n$, we have $(\widetilde A_j^{(\lambda )})_N=\widehat \mu \, (2\pi N_1
^{(\lambda )})\, (A_j^{(\lambda )})_N$ if $N_2=0$ and $(\widetilde A_j^{(\lambda )})
_N=0$ if $N_2\neq 0$. (Here, $(A_j^{(\lambda )})_N$ are the Fourier coefficients of
$A_j^{(\lambda )}$, $N\in \Lambda ^*$.)

Let $\Phi ^{(s,\lambda )}:{\mathbb R}^n\to {\mathbb C}$, $s=1,2$, be periodic 
trigonometric polynomials with the period lattice $\Lambda $ and the Fourier 
coefficients $\Phi ^{(1,\lambda )}_N=\Phi ^{(2,\lambda )}_N=0$ if $N_1^{(\lambda )}=
N_2=0$ and
$$
\Phi ^{(1,\lambda )}_N=\bigl( \, 2\pi i\, \bigl( (N_1^{(\lambda )})^2+N_2^2\bigr) 
\bigr) ^{-1}\bigl( N_1^{(\lambda )}(A_1^{(\lambda )}-\widetilde A_1^{(\lambda )})_N+
N_2\, (A_2-\widetilde A_2)_N\bigr) ,
$$ $$
\Phi ^{(2,\lambda )}_N=-\bigl( \, 2\pi i\, \bigl( (N_1^{(\lambda )})^2+N_2^2\bigr) 
\bigr) ^{-1}\bigl( N_2\, (A_1^{(\lambda )}-\widetilde A_1^{(\lambda )})_N-N_1
^{(\lambda )}(A_2-\widetilde A_2)_N\bigr) 
$$
otherwise. We have
$$
\frac {\partial \Phi ^{(1,\lambda )}}{\partial x_1^{(\lambda )}}-\frac {\partial 
\Phi ^{(2,\lambda )}}{\partial x_2}=A_1^{(\lambda )}-\widetilde A_1^{(\lambda )},
\qquad \frac {\partial \Phi ^{(1,\lambda )}}{\partial x_2}+\frac {\partial \Phi 
^{(2,\lambda )}}{\partial x_1^{(\lambda )}}=A_2-\widetilde A_2\, .
$$
\vskip 0.2cm

{\bf Lemma 1.} {\it There is a universal constant $C>0$ such that
$$
\| \Phi ^{(s,\lambda )}\| _{L^{\infty }({\mathbb R}^n)}\leq C\, \| \mu \| \, \max \,
\{ |\gamma |,h^{-1}\} \, \| A\| _{L^{\infty }({\mathbb R}^n;{\mathbb C}^n)}\, ,\quad
s=1,2.
$$}
\vskip 0.2cm

{\bf Proof.} Let $\eta (.)\in C^{\infty }({\mathbb R};{\mathbb R})$, $\eta (\tau )=0$
for $\tau \leq \pi $, $0\leq \eta (\tau )\leq 1$ for $\pi <\tau \leq 2\pi $, and
$\eta (\tau )=1$ for $\tau >2\pi $. For $x,y\in {\mathbb R}$ (and $x^2+y^2>0$), we
set
$$
G(x,y)=\frac x{x^2+y^2}\, \int\limits_0^{+\infty }\frac {\partial \eta (\tau )}
{\partial \tau }\, J_0\bigl( \tau \sqrt {x^2+y^2}\, \bigr) \, d\tau \, ,
$$
where $J_0(.)$ is the Bessel function of the first kind of order zero; $G(.,.)\in
L^q({\mathbb R}^2)$, $q\in [1,2)$. We write $G_1(t;x,y)=t^{-1}G(t^{-1}x,t^{-1}y)$,
$t>0$, and $G_2(t;x,y)=G_1(t;y,x)$; $\| G_s(t;.,.)\| _{L^1({\mathbb R}^2)}=t\, \|
G(.,.)\| _{L^1({\mathbb R}^2)}\, $, $s=1,2$. For arbitrary continuous periodic 
functions ${\mathcal F}:{\mathbb R}^n\to {\mathbb C}$ with the period lattice 
$\Lambda $, we set
$$
\bigl( {\mathcal F}*_{\lambda }G_s(t;.,.)\bigr) (x)=\iint\limits_{{\mathbb R}^2}
G_s(t;\xi _1,\xi _2)\, {\mathcal F}(x-\xi _1\widetilde e^{\lambda }-\xi _2e)\, d\xi _1
d\xi _2\, ,\quad x\in {\mathbb R}^n.
$$
In this case, $( {\mathcal F}*_{\lambda }G_s(t;.,.))_N=0$ if $N_1^{(\lambda )}=N_2=0$
and
$$
\bigl( {\mathcal F}*_{\lambda }G_s(t;.,.)\bigr) _N=-\, \frac {iN_s^{(\lambda )}}
{(N_1^{(\lambda )})^2+N_2^2}\, \eta \biggl( 2\pi t\sqrt {(N_1^{(\lambda )})^2+N_2^2}
\, \biggr) \, {\mathcal F}_N
$$
otherwise, $s=1,2$. Let $t=\max \, \{ |\gamma |,h^{-1}\} $. Because $(A-\widetilde 
A)_N=0$ for $N_2=0$, $|N_1^{(\lambda )}|\leq h$, and $|N_2|=|\gamma |^{-1}|(N,\gamma
)|\geq |\gamma |^{-1}$ for $N_2\neq 0$, we have
$$
2\pi \Phi ^{(1,\lambda )}=(A_1^{(\lambda )}-\widetilde A_1^{(\lambda )})*_{\lambda }
G_1(t;.,.)+(A_2-\widetilde A_2)*_{\lambda }G_2(t;.,.),
$$ $$
2\pi \Phi ^{(2,\lambda )}=-\, (A_1^{(\lambda )}-\widetilde A_1^{(\lambda )})*_{\lambda }
G_2(t;.,.)+(A_2-\widetilde A_2)*_{\lambda }G_1(t;.,.).
$$
Using the inequalities $\| \widetilde A\| _{L^{\infty }({\mathbb R}^n;{\mathbb C}^n)}
\leq \| \mu \| \, \| A\| _{L^{\infty }({\mathbb R}^n;{\mathbb C}^n)}$ and $\| \mu \|
\geq 1$, and taking the constant $C=2\pi ^{-1}\| G(.,.)\| _{L^1({\mathbb R}^2)}\, $,
we complete the proof of the lemma. \hfill $\square $
\vskip 0.2cm

We introduce the notation 
$$ 
\widehat D_0^{(\lambda )}=\biggl( -i\, \frac {\partial }{\partial x_1^{(\lambda )}}+
k_1^{(\lambda )}\biggr) \, \widehat \alpha _1^{(\lambda )}+\biggl( -i\, \frac
{\partial }{\partial x_2}+k_2+i\varkappa \biggr) \, \widehat \alpha _2\, ,
$$ $$
\widehat D^{(\lambda )} =\widehat D_0^{(\lambda )}-\widetilde A_1^{(\lambda )}
\widehat \alpha _1^{(\lambda )}-\widetilde A_2\widehat \alpha _2\, ,
$$ $$
\widehat D^{(\lambda )}(k+i\varkappa e)=e^{-i\widehat \alpha _1^{(\lambda )}
\widehat \alpha _2\, \Phi ^{(2,\lambda )}}\, e^{\, i\Phi ^{(1,\lambda )}}\, \widehat 
D^{(\lambda )}\, e^{-i\Phi ^{(1,\lambda )}}\, e^{-i\widehat \alpha _1^{(\lambda )}
\widehat \alpha _2\, \Phi ^{(2,\lambda )}},
$$ $$
\widehat {\mathcal V}^{(\lambda )}=\widehat V^{(0)}+\widehat V^{(1)}+\sum\limits
_{j=3}^n\biggl( -i\, \frac {\partial }{\partial x_j^{(\lambda )}}+k_j^{(\lambda )}
-A_j^{(\lambda )} \biggr) \, \widehat \alpha _j^{(\lambda )},
$$ $$
\widehat D(k+i\varkappa e)=\widehat D^{(\lambda )}(k+i\varkappa e)+\widehat 
{\mathcal V}^{(\lambda )}.
$$

If $N\in {\mathcal K}_{\beta }^{\lambda }\, $, then $|\widetilde e(k+2\pi N)-
\widetilde e^{\lambda }|<\rho $ and therefore
$$
|k+2\pi N-(k_2+2\pi N_2)e-\varkappa \widetilde e^{\lambda }|<\beta +\rho \varkappa 
\, .
$$
It follows that
$$
\bigg| \, \sum\limits_{j=3}^n\, (k_j^{(\lambda )}+2\pi N_j^{(\lambda )})\, {\mathcal
E}_j^{(\lambda )}\, \biggr| <\beta +\rho \varkappa \, ,\quad |k_1^{(\lambda )}+2\pi 
N_1^{(\lambda )}-\varkappa |<\beta +\rho \varkappa \, ,  \eqno (9)
$$
and
$$
\| \widehat {\mathcal V}^{(\lambda )}\phi ^{\, {\mathcal K}_{\beta }^{\lambda }}\|
\leq \bigl( \beta +(c_6+2)R+W\bigr) \, \| \phi ^{\, {\mathcal K}_{\beta }^{\lambda }}
\| \, .
$$
We use the brief notation $\widehat P^{\, \pm }_{\lambda }=\widehat P^{\, \pm }
_{\widetilde e^{\lambda }}=(1/2)(\widehat I\pm i\widehat \alpha _1^{(\lambda )}
\widehat \alpha _2)$. We set $\chi ^{(\lambda )}=e^{-i\Phi ^{(1,\lambda )}}\, 
e^{-i\widehat \alpha _1^{(\lambda )}\widehat \alpha _2\, \Phi ^{(2,\lambda )}}\phi 
^{\, {\mathcal K}_{\beta }^{\lambda }}$. The relation
$$
\widehat D_0^{(\lambda )}\widehat P^{\, \pm }_{\lambda }\chi ^{(\lambda )}=
\sum\limits_{N\, \in \, \Lambda ^*}\bigl( k_2+2\pi N_2+i(k_1^{(\lambda )}+2\pi N_1
^{(\lambda )}\, ))\bigr) \, \widehat \alpha _2\widehat P^{\, \pm }_{\lambda }\chi _N
^{(\lambda )}\, e^{\, 2\pi i\, (N,x)}  \eqno (10)
$$
holds.

We write ${\mathcal O}^{(\lambda )}(\tau )=\{ N\in \Lambda ^*:|k_1^{(\lambda )}+2\pi 
N_1^{(\lambda )}-\varkappa |<2\tau \} $, $\tau >0$. Inequalities (9) imply that
there is a constant
$$
c_7=c_7\, (\widetilde \theta ,\theta ,W_0,R,\beta )>\frac 12\, (\beta +(c_6+2)R)
$$
such that (for all $\lambda $)
$$
\biggl\| \sum\limits_{N\, \in \, \Lambda ^*\, \backslash \, {\mathcal O}^{(\lambda )}
(c_7)}\widehat P^{\, +}_{\lambda }\chi _N^{(\lambda )}\, e^{\, 2\pi i\, (N,x)}\,
\biggr\| \leq \frac 12\, \bigl\| \widehat P^{\, +}_{\lambda }\chi ^{(\lambda )}
\bigr\| \, .  \eqno (11)
$$

In what follows, we assume that $\varkappa _0\geq c_7\, $. As a consequence of (10)
and (11), we obtain
$$
\| \widehat D_0^{(\lambda )}\widehat P^{\, +}_{\lambda }\chi ^{(\lambda )}\| \geq
v^{\, 1/2}(K)\, \biggl( \sum\limits_{N\, \in \, {\mathcal O}^{(\lambda )}(c_7)}
|\varkappa +(k_1^{(\lambda )}+2\pi N_1^{(\lambda )})|^2\, \| \widehat P^{\, +}
_{\lambda }\chi _N^{(\lambda )}\| ^2\biggr) ^{1/2}\geq
$$ $$
2\, (\varkappa -c_7)\, \biggl\| \sum\limits_{N\, \in \, {\mathcal O}^{(\lambda )}
(c_7)}\widehat P^{\, +}_{\lambda }\chi _N^{(\lambda )}\, e^{\, 2\pi i\, (N,x)}\,
\biggr\| \geq (\varkappa -c_7)\, \| \widehat P^{\, +}_{\lambda }\chi ^{(\lambda )}
\| \, .
$$
On the other hand, we have $|k_2+2\pi N_2|\geq \pi |\gamma |^{-1}$. Condition (6)
implies that
$$
\| \, \widetilde A_1^{(\lambda )}\widehat \alpha _1^{(\lambda )}+\widetilde A_2
\widehat \alpha _2\, \| _{L^{\infty }({\mathbb R}^n;{\mathcal L}_M)}\, \leq \, 
\widetilde \theta \pi |\gamma |^{-1},
$$
and therefore (see (10))
$$
\| \widehat D^{(\lambda )}\widehat P^{\, -}_{\lambda }\chi ^{(\lambda )}\| \geq
\| \widehat D_0^{(\lambda )}\widehat P^{\, -}_{\lambda }\chi ^{(\lambda )}\| -
\widetilde \theta \pi |\gamma |^{-1}\| \widehat P^{\, -}_{\lambda }\chi ^{(\lambda 
)}\| \geq (1-\widetilde \theta \, )\pi |\gamma |^{-1}\| \widehat P^{\, -}_{\lambda }
\chi ^{(\lambda )}\| \, .
$$
The operators $\widehat P^{\, \pm }_{\lambda }$ commute with the operators $e^{\,
\pm i\Phi ^{(1,\lambda )}}$, $e^{-i\widehat \alpha _1^{(\lambda )}\widehat \alpha _2\, 
\Phi ^{(2,\lambda )}}$, and $\widehat {\mathcal V}^{(\lambda )}$, and we have 
$\widehat P^{\, \pm }_{\lambda }\widehat D^{(\lambda )}=\widehat D^{(\lambda )}
\widehat P^{\, \mp }_{\lambda }\, $. Consequently,
$$
\widehat P^{\, \pm }_{\lambda }\widehat D(k+i\varkappa e)=\widehat D^{(\lambda )}(k+
i\varkappa e)\widehat P^{\, \mp }_{\lambda }+\widehat {\mathcal V}^{(\lambda )}
\widehat P^{\, \pm }_{\lambda }\, .
$$
Using the above estimates and also the inequality
$$
\| \, e^{\, \pm \, i\Phi ^{(1,\lambda )}}\, e^{\, i\widehat \alpha _1^{(\lambda )}
\widehat \alpha _2\, \Phi ^{(2,\lambda )}}\, \| _{L^{\infty }({\mathbb R}^n;
{\mathcal L}_M)}\, \leq \, c_5^{-1/2}\, (A),
$$
we derive
$$
\| \widehat P^{\, +}_{\lambda }\widehat D(k+i\varkappa e)\phi ^{\, {\mathcal K}
_{\beta }^{\lambda }}\| \geq (1-\widetilde \theta \, )\pi |\gamma |^{-1}c_5(A)\,
\| \widehat P^{\, -}_{\lambda }\phi ^{\, {\mathcal K}_{\beta }^{\lambda }}\| -
\| \widehat {\mathcal V}^{(\lambda )}\widehat P^{\, +}_{\lambda }\phi ^{\, {\mathcal 
K}_{\beta }^{\lambda }}\| \, ,  \eqno (12)
$$ $$
\| \widehat P^{\, -}_{\lambda }\widehat D(k+i\varkappa e)\phi ^{\, {\mathcal K}
_{\beta }^{\lambda }}\| \, \geq  \eqno (13)
$$ $$
(\varkappa -c_7-\pi |\gamma |^{-1})\, c_5(A)\,
\| \widehat P^{\, +}_{\lambda }\phi ^{\, {\mathcal K}_{\beta }^{\lambda }}\| -
\| \widehat {\mathcal V}^{(\lambda )}\widehat P^{\, -}_{\lambda }\phi ^{\, {\mathcal 
K}_{\beta }^{\lambda }}\| \, .  
$$
Let
$$
\sigma =\theta _2^2\, \theta _3^{-2}-1,
$$ $$
\widetilde a=\min \, \{ 1, \sqrt {\sigma }\, a, (1-\widetilde \theta -\theta _1)\, \pi
|\gamma |^{-1}c_5^{\, \prime }\, (\beta +(c_6+2)R+W_0)^{-1}\} ,
$$ 
\vskip 0.1cm
$$
b^{\, \prime \prime }=\min \, \left\{ 
\begin{array}{llll}
1\, , \\ [0.2cm]
\sqrt {\sigma }\, a\, , \\ [0.2cm]
(1-\widetilde \theta -\theta _1)\, \pi |\gamma |^{-1}\, c_5^{\, \prime }\, 
(c_6+2)^{-1}(1+W_0)^{-1}\, , \\ [0.2cm]
2\, (\theta _1-\theta _2)\, \pi |\gamma |^{-1}(c_6+2)^{-1}\, .
\end{array}
\right.
$$
\vskip 0.2cm
Since $(\beta +R)^{-1}\, b^{\, \prime \prime }<\widetilde a$, we can pick a number 
$\widetilde a^{\, \prime }$ such that $(\beta +R)^{-1}\, b^{\, \prime \prime }\leq 
\widetilde a^{\, \prime }<\widetilde a$. For an adequate choice of the number 
$\varkappa _0$ (and for $\varkappa \geq \varkappa _0$), inequalities (12) and (13)
imply the estimate
$$
\| \widehat P^{\, +}_{\lambda }\widehat D(k+i\varkappa e)\phi ^{\, {\mathcal K}
_{\beta }^{\lambda }}\| +\widetilde a\, \| \widehat P^{\, -}_{\lambda }\widehat 
D(k+i\varkappa e)\phi ^{\, {\mathcal K}_{\beta }^{\lambda }}\| \, \geq 
$$ $$
c_5(A)\, \bigl( \, \theta _1 \pi |\gamma |^{-1}\| \widehat P^{\, -}_{\lambda }\phi ^{\, 
{\mathcal K}_{\beta }^{\lambda }}\| +\widetilde a^{\, \prime }\varkappa \, \| 
\widehat P^{\, +}_{\lambda }\phi ^{\, {\mathcal K}_{\beta }^{\lambda }}\| \, \bigr) .
$$
For all $\widetilde e\in \widehat S\Omega _{\lambda }\subset S_{n-2}(e)$, we have
$$
\bigl\| \, (\widehat P^{\, \pm }-\widehat P^{\, \pm }_{\lambda })\phi ^{\, {\mathcal 
K}_{\beta }^{\lambda }}\, \bigr\| \leq \frac 12\, |\widetilde e-\widetilde e^{\lambda
}|\, \|\phi ^{\, {\mathcal K}_{\beta }^{\lambda }}\| \leq \frac {\rho }2\, \|\phi 
^{\, {\mathcal K}_{\beta }^{\lambda }}\| \, .  \eqno (14)
$$
If $(\widehat D(k+i\varkappa e)\phi ^{\, {\mathcal K}_{\beta }^{\lambda }})_N\neq 0$
for some $N\in \Lambda ^*$, then $k+2\pi N\notin {\mathcal P}(e)$ and $|\widetilde
e(k+2\pi N)-\widetilde e^{\lambda }|<\rho +2R/\varkappa =\rho ^{\, \prime }$. 
Therefore,
$$
\bigl\| \, (\widehat P^{\, \pm }-\widehat P^{\, \pm }_{\lambda })\widehat D(k+i
\varkappa e)\phi ^{\, {\mathcal K}_{\beta }^{\lambda }}\, \bigr\| \leq \frac {\rho
^{\, \prime }}2\, \bigl\| \, \widehat D(k+i\varkappa e)\phi ^{\, {\mathcal K}_{\beta }
^{\lambda }}\, \bigr\| \, .
$$
Consequently,
$$
\| \widehat P^{\, +}_{\lambda }\widehat D(k+i\varkappa e)\phi ^{\, {\mathcal K}
_{\beta }^{\lambda }}\| +\widetilde a\, \| \widehat P^{\, -}_{\lambda }\widehat 
D(k+i\varkappa e)\phi ^{\, {\mathcal K}_{\beta }^{\lambda }}\| \, \leq   \eqno (15)
$$ $$
(1+\rho ^{\, \prime }\, \widetilde a^{-1})\, \bigl( \, \| \widehat P^{\, +}\widehat 
D(k+i\varkappa e)\phi ^{\, {\mathcal K}_{\beta }^{\lambda }}\| +\widetilde a\, \| 
\widehat P^{\, -}\widehat D(k+i\varkappa e)\phi ^{\, {\mathcal K}_{\beta }^{\lambda 
}}\| \, \bigr) . 
$$
Since $(\beta +R)^{-1}\, b^{\, \prime \prime }\leq \widetilde a^{\, \prime }$ and
$(c_6+2)\, b^{\, \prime \prime }\leq 2\, (\theta _1-\theta _2)\, \pi |\gamma |^{-1}$,
for an adequately chosen number $\varkappa _0$ (and for $\varkappa \geq \varkappa 
_0$) inequality (14) implies that
$$
\theta _1\, \frac {\pi }{|\gamma |}\, \| \widehat P^{\, -}_{\lambda }\phi ^{\, 
{\mathcal K}_{\beta }^{\lambda }}\| +\frac {b^{\, \prime \prime }\varkappa }{\beta +
R}\, \| \widehat P^{\, +}_{\lambda }\phi ^{\, {\mathcal K}_{\beta }^{\lambda }}\|
\, \geq  \eqno (16)
$$ $$
(1+\rho ^{\, \prime }\, \widetilde a^{-1})\, \biggl( \theta _2\, \frac {\pi }
{|\gamma |}\, \| \widehat P^{\, -}\phi ^{\, {\mathcal K}_{\beta }^{\lambda }}\| +
\frac {b^{\, \prime \prime }\varkappa }{2\, (\beta +R)}\, \| \widehat P^{\, +}\phi 
^{\, {\mathcal K}_{\beta }^{\lambda }}\| \biggr) .
$$
From (15) and (16), it follows that
$$
\| \widehat P^{\, +}\widehat D(k+i\varkappa e)\phi ^{\, {\mathcal K}_{\beta }
^{\lambda }}\| +\widetilde a\, \| \widehat P^{\, -}\widehat D(k+i\varkappa e)\phi 
^{\, {\mathcal K}_{\beta }^{\lambda }}\| \, \geq 
$$ $$
c_5(A)\, \biggl( \theta _2\, \frac {\pi }{|\gamma |}\, \| \widehat P^{\, -}\phi 
^{\, {\mathcal K}_{\beta }^{\lambda }}\| +\frac {b^{\, \prime \prime }\varkappa }{2\, 
(\beta +R)}\, \| \widehat P^{\, +}\phi ^{\, {\mathcal K}_{\beta }^{\lambda }}\| 
\biggr) .
$$
We write $b^{\, \prime }=(1/2)(1+\sigma )^{-1/2}\, b^{\, \prime \prime }$. Then
$$
\| \widehat P^{\, +}\widehat D(k+i\varkappa e)\phi ^{\, {\mathcal K}_{\beta }
^{\lambda }}\| ^2+a^2\, \| \widehat P^{\, -}\widehat D(k+i\varkappa e)\phi 
^{\, {\mathcal K}_{\beta }^{\lambda }}\| ^2\, \geq  \eqno (17)
$$ $$
(1+\sigma )^{-1}\, \bigl( \, \| \widehat P^{\, +}\widehat D(k+i\varkappa e)\phi ^{\, 
{\mathcal K}_{\beta }^{\lambda }}\| +\widetilde a\, \| \widehat P^{\, -}\widehat D(k
+i\varkappa e)\phi ^{\, {\mathcal K}_{\beta }^{\lambda }}\| \, \bigr) ^2 \, \geq
$$ $$
c_5^2(A)\, \biggl( \biggl( \theta _3\, \frac {\pi }{|\gamma |}\, \biggr) ^2 \| \widehat 
P^{\, -}\phi ^{\, {\mathcal K}_{\beta }^{\lambda }}\| ^2+\biggl( \frac {b^{\, \prime }
\varkappa }{2\, (\beta +R)}\biggr) ^2\, \| \widehat P^{\, +}\phi ^{\, {\mathcal 
K}_{\beta }^{\lambda }}\| ^2\biggr) . 
$$

If $N\in \Lambda ^*$ and $\lambda _1\neq \lambda _2\, $, then either $2\pi |N-N^{\,
\prime }|>R$ for all $N^{\, \prime }\in {\mathcal K}_{\beta }^{\lambda _1}$ or
$2\pi |N-N^{\, \prime \prime }|>R$ for all $N^{\, \prime \prime }\in {\mathcal K}
_{\beta }^{\lambda _2}$. Therefore,
$$
\widehat V\phi ^{\, \bigcup\limits_{\lambda }\, {\mathcal K}_{\beta }^{\lambda }}=
\sum\limits_{\lambda }\widehat V\phi ^{\, {\mathcal K}_{\beta }^{\lambda }}\, ,\qquad
\widehat D(k+i\varkappa e)\, \phi ^{\, \bigcup\limits_{\lambda }\, {\mathcal K}_{\beta 
}^{\lambda }}=\sum\limits_{\lambda }\widehat D(k+i\varkappa e)\, \phi ^{\, {\mathcal 
K}_{\beta }^{\lambda }}\, .
$$
If $N\in \bigcup\limits_{\lambda }\, {\mathcal K}_{\beta }^{\lambda }\, $, then
$$
(\widehat D(k+i\varkappa e)\, \phi )_N=\biggl( \widehat D(k+i\varkappa e)\, \phi ^{\, 
\bigcup\limits_{\lambda }\, {\mathcal K}_{\beta }^{\lambda }}\, \biggr) _N+\biggl(
\widehat V\, \phi ^{\, {\mathcal K}_{\beta }\backslash \, \bigcup\limits_{\lambda }\, 
{\mathcal K}_{\beta }^{\lambda }}\, \biggr) _N\, . 
$$
If $N\in \Lambda ^*\, \backslash \bigcup\limits_{\lambda }\, {\mathcal K}_{\beta }
^{\lambda }\, $, then
$$
\biggl( \widehat D(k+i\varkappa e)\, \phi ^{\, \bigcup\limits_{\lambda }\, {\mathcal 
K}_{\beta }^{\lambda }}\, \biggr) _N=\biggl( \widehat V\, \phi ^{\, \bigcup\limits
_{\lambda }\, (\, {\mathcal K}_{\beta }^{\lambda }\, \backslash \, \widetilde 
{\mathcal K}_{\beta }^{\lambda }\, )}\biggr) _N\, . 
$$
These relations (for each of the signs) imply the estimates
$$
\| \widehat P^{\, \pm }\widehat D(k+i\varkappa e)\, \phi \| ^2 \, \geq  \eqno (18)
$$ $$
v(K)\, \sum\limits_{N\, \in \, \bigcup\limits_{\lambda }\, {\mathcal K}_{\beta }
^{\lambda }}\ \biggl\| \biggl( \widehat P^{\, \pm }\widehat D(k+i\varkappa e)\, \phi 
^{\, \bigcup\limits_{\lambda }\, {\mathcal K}_{\beta }^{\lambda }}\, \biggr) _N+
\biggl( \widehat P^{\, \pm }\widehat V\, \phi ^{\, {\mathcal K}_{\beta }\backslash \, 
\bigcup\limits_{\lambda }\, {\mathcal K}_{\beta }^{\lambda }}\, \biggr) _N\biggr\|
\, \geq
$$ $$
(1-\delta )\, \biggl\| \, \widehat P^{\, \pm }\, \widehat D(k+i\varkappa e)\, \phi 
^{\, \bigcup\limits_{\lambda }\, {\mathcal K}_{\beta }^{\lambda }}\, \biggr\| ^2\, -
$$ $$
(1-\delta )\, \biggl\| \, \widehat P^{\ \Lambda ^*\, \backslash \, \bigcup\limits
_{\lambda }\, {\mathcal K}_{\beta }^{\lambda }}\, \widehat P^{\, \pm }\, \widehat V\, 
\phi ^{\, \bigcup\limits_{\lambda }\, (\, {\mathcal K}_{\beta }^{\lambda }\, 
\backslash \, \widetilde {\mathcal K}_{\beta }^{\lambda }\, )}\, \biggr\| ^2\, -
$$ $$
(1-\delta )\, \delta ^{-1}\, \biggl\| \, \widehat P^{\ \bigcup\limits_{\lambda }\, 
{\mathcal K}_{\beta }^{\lambda }}\, \widehat P^{\, \pm }\, \widehat V\, \phi ^{\, 
{\mathcal K}_{\beta }\backslash \, \bigcup\limits_{\lambda }\, {\mathcal K}_{\beta }
^{\lambda }}\, \biggr\| ^2\, \geq
$$ $$
(1-\delta )\, \biggl\| \, \widehat P^{\, \pm }\widehat D(k+i\varkappa e)\, \phi ^{\, 
\bigcup\limits_{\lambda }\, {\mathcal K}_{\beta }^{\lambda }}\, \biggr\| ^2-
(1-\delta ^2)\, \delta ^{-1}W^2\, \biggl\| \, \phi ^{\, {\mathcal K}_{\beta }\backslash 
\, \bigcup\limits_{\lambda }\, {\mathcal K}_{\beta }^{\lambda }}\, \biggr\| ^2\, \geq 
$$ $$
(1-\delta )\, \biggl\| \, \widehat P^{\, \pm }\widehat D(k+i\varkappa e)\, \phi ^{\, 
\bigcup\limits_{\lambda }\, {\mathcal K}_{\beta }^{\lambda }}\, \biggr\| ^2-
\widetilde \varepsilon \, \delta ^{-1}W^2\, \| \phi \| ^2.
$$
We set $b=(1/4)\sqrt {(1-\delta )(1-\widetilde \varepsilon \, )}\, b^{\, \prime }$. For
$\varkappa _0\, $, we assume that $3(c_5^{\, \prime })^2b^{\, 2}\varkappa _0^2\geq
8\, \widetilde \varepsilon \, \delta ^{-1}W_0^2R^2$. Then for $\varkappa \geq
\varkappa _0\, $, from (17) and (18) (in view of (8) and the constraint $a\leq 1$) we 
obtain the inequalities
$$
\| \widehat P^{\, +}\widehat D(k+i\varkappa e)\phi \| ^2+a^2\, \| \widehat P^{\, -}
\widehat D(k+i\varkappa e)\phi \| ^2\, \geq 
$$ $$
(1-\delta )\, c_5^2(A)\, \sum\limits_{\lambda }\, \biggl( \biggl( \theta _3\, \frac 
{\pi }{|\gamma |}\, \biggr) ^2 \| \widehat P^{\, -}\phi ^{\, {\mathcal K}_{\beta }
^{\lambda }}\| ^2+\biggl( \frac {b^{\, \prime }\varkappa }{2\, (\beta +R)}\biggr) 
^2\, \| \widehat P^{\, +}\phi ^{\, {\mathcal K}_{\beta }^{\lambda }}\| ^2\biggr) \, -
$$ $$
\frac 2{\delta }\, W^2\widetilde \varepsilon \, \| \phi \| ^2\, \geq
$$ $$
(1-\widetilde \varepsilon \, )\, c_5^2(A)\, \biggl( \biggl( \theta _4\, \frac {\pi }
{|\gamma |}\, \biggr) ^2 \| \widehat P^{\, -}\phi \| ^2+(1-\delta )\, \biggl( \frac 
{b^{\, \prime }\varkappa }{2\, (\beta +R)}\biggr) ^2\, \| \widehat P^{\, +}\phi \| 
^2\biggr) \, -
$$ $$
\frac 2{\delta }\, W^2\widetilde \varepsilon \, (\| \widehat P^{\, -}\phi 
\| ^2+\| \widehat P^{\, +}\phi \| ^2)\, \geq
$$ $$
c_5^2(A)\, \biggl( \biggl( \theta \, \frac {\pi }{|\gamma |}\, \biggr) ^2 \| \widehat 
P^{\, -}\phi \| ^2+(1-\delta )\, \biggl( \frac {b\varkappa }{\beta +R}\biggr) ^2\, \| 
\widehat P^{\, +}\phi \| ^2\biggr) .
$$
Theorem 5 is proved.
\vskip 0.2cm

{\bf 3.} The following theorems are a consequence of Theorem 5. The proof of Theorem
6 is based on applying the relation
$$
\widehat P^{\, \pm }(k)\widehat D_0(k+i\varkappa e)=\widehat D_0(k+i\varkappa e)
\widehat P^{\, \mp }(k)
$$
and on selecting an arbitrarily small number $a\in (0,1]$. The proof of Theorem 7
essentially uses the arbitrariness in the choice of the number $\beta >0$ (see
below). Theorem 6 is used to prove the absolute continuity of the spectrum of a
periodic Schr\"odinger operator.
\vskip 0.2cm

{\bf Theorem 6.} {\it Let $\widetilde \theta \in [0,1)$, $W_0\geq 0$, $R\geq 1$, and
$\beta >0$ $\mathrm ($for a fixed vector $\gamma \in \Lambda \backslash \{ 0\} $ and
a fixed measure $\mu \in {\mathcal M}_h\, $, $h>0$$\mathrm ;$ $e=|\gamma |^{-1}\gamma 
$$\mathrm )$. Then there are numbers $c_8=c_8\, (\widetilde \theta ,W_0)>0$ and 
$\varkappa _0=\varkappa _0\, (\widetilde \theta ,W_0,R,\beta )>4\beta +5R$ such that 
for all vectors $k\in {\mathbb R}^n$ with $(k,\gamma )=\pi $, all $\varkappa \geq 
\varkappa _0\, $, all continuous periodic functions $\widehat V^{(s)}:{\mathbb R}^n\to
{\mathcal L}^{(s)}_M$, $s=0,1$, and $A:{\mathbb R}^n\to {\mathbb C}^n$ $\mathrm 
($with the period lattice $\Lambda \subset {\mathbb R}^n$, $n\geq 3$$\mathrm )$ for
which $A_0=0$ and conditions $\mathrm {(5)}$ -- $\mathrm {(7)}$ are satisfied,
and all vector functions $\phi \in {\mathcal H}({\mathcal K}_{\beta }(k;\varkappa ))$,
the inequality
$$
\| \widehat D^2(k+i\varkappa e)\phi \| \geq \frac {c_8\, \varkappa }{\beta +R}\,
\| \phi \|
$$
holds.}
\vskip 0.2cm

{\bf Theorem 7.} {\it Let $\widetilde \theta \in [0,1)$, $\theta \in (0,1-\widetilde 
\theta )$, $W_0\geq 0$, $R\geq 1$, and $\delta \in (0,1]$ $\mathrm ($for a fixed vector 
$\gamma \in \Lambda \backslash \{ 0\} $ and a fixed measure $\mu \in {\mathcal M}_h\, 
$, $h>0$$\mathrm ;$ $e=|\gamma |^{-1}\gamma $$\mathrm )$. Then there are numbers 
${\mathcal D}={\mathcal D}\, (\theta ,W_0,\delta )\geq 1$ and $\varkappa _0=\varkappa 
_0\, (\widetilde \theta ,\theta ,W_0,R,\delta )>(4{\mathcal D}+1)R$ such that for all
vectors $k\in {\mathbb R}^n$ with $(k,\gamma )=\pi $, all $\varkappa \geq \varkappa
_0\, $, all continuous periodic functions $\widehat V^{(s)}:{\mathbb R}^n\to
{\mathcal L}^{(s)}_M$, $s=0,1$, and $A:{\mathbb R}^n\to {\mathbb C}^n$ $\mathrm 
($with the period lattice $\Lambda \subset {\mathbb R}^n$, $n\geq 3$$\mathrm )$ for
which $A_0=0$ and conditions $\mathrm {(5)}$ -- $\mathrm {(7)}$ are satisfied,
and all vector functions $\phi \in \widetilde H^1(K;{\mathbb C}^M)$, the inequality
$$
\| \widehat D(k+i\varkappa e)\phi \| ^2 \, \geq 
$$ $$
(1-\delta )\, \biggl( c_5^2(A)\, 
\biggl( \theta \, \frac {\pi }{|\gamma |}\, \biggr) ^2 \bigl\| \phi ^{\, {\mathcal K}
_{{\mathcal D}R}}\bigr\| ^2+v(K)\, \sum\limits_{N\, \in \, \Lambda ^*\backslash \,
{\mathcal K}_{{\mathcal D}R}}G_N^2(k;\varkappa )\, \| \phi _N\| ^2\biggr)
$$
holds.}
\vskip 0.2cm

{\bf Proof of Theorem 2.} Let $\theta <\theta ^{\, \prime }<1-\widetilde \theta $,
$\widehat V^{(s)}_{\nu }:{\mathbb R}^n\to {\mathcal L}^{(s)}_M\, $, $s=0,1$, and
$A_{\nu }:{\mathbb R}^n\to {\mathbb C}^n$, $\nu \in {\mathbb N}$, be sequences of
trigonometric polynomials with the period lattice $\Lambda $ that uniformly converge
as $\nu \to +\infty $ to the functions $\widehat V^{(s)}$ and $A$, let $(A_{\nu })
_0=0$ for all $\nu \in {\mathbb N}$, and let $\widehat V_{\nu }=\widehat V_{\nu }
^{(0)}+\widehat V_{\nu }^{(1)}-\sum\limits_{j=1}^n(A_{\nu })_j\widehat \alpha _j\, $.
From Theorem 7 (because $G_N(k;\varkappa )\geq \pi |\gamma |^{-1}$, $N\in \Lambda
^*$) it follows that for all sufficiently large $\nu $, there are numbers $\varkappa
_0^{(\nu )}>0$ such that for all $k\in {\mathbb R}^n$ with $(k,\gamma )=\pi $, all 
$\varkappa \geq \varkappa _0^{(\nu )}$, and all vector functions $\phi \in \widetilde 
H^1(K;{\mathbb C}^M)$, the inequality
$$
\| (\widehat D_0(k+i\varkappa e)+\widehat V_{\nu })\phi \| \geq c_5(A_{\nu })\,
\theta ^{\, \prime }\pi |\gamma |^{-1}\| \phi \|
$$
is valid. For a sufficiently large index $\nu $ (and for $\varkappa \geq \varkappa
_0^{(\nu )}$), it follows that the desired inequality holds. Theorem 2 is proved.
\vskip 0.2cm

{\bf Theorem 8.} {\it Let $\widehat V^{(s)}:{\mathbb R}^n\to {\mathcal L}^{(s)}_M$, 
$s=0,1$, and $A:{\mathbb R}^n\to {\mathbb C}^n$ be continuous periodic functions 
with the period lattice $\Lambda \subset {\mathbb R}^n$, $n\geq 3$. If $A_0=0$ and 
condition $\mathrm {(6)}$ with $\widetilde \theta \in [0,1)$ is satisfied for a 
vector $\gamma \in \Lambda \backslash \{ 0\} $ $\mathrm ($$e=|\gamma |^{-1}\gamma 
$$\mathrm )$ and a measure $\mu \in {\mathcal M}_h\, $, $h>0$, then for any
$\delta \in (0,1]\, $, there are numbers $\beta =\beta \, (\gamma ,h,\mu ;\widehat
V,\delta )>0$ and $\varkappa _0=\varkappa _0\, (\gamma ,h,\mu ;\widehat V,\delta )>0$
such that for all $k\in {\mathbb R}^n$ with $(k,\gamma )=\pi $, all $\varkappa \geq 
\varkappa _0\, $,  and all vector functions $\phi \in \widetilde H^1(K;{\mathbb C}
^M)$, the inequality
$$
\| \widehat D(k+i\varkappa e)\phi \| ^2 \, \geq \, 
(1-\delta )\, \biggl( c_5^2(A;\gamma ,h,\mu )\, (1-\widetilde \theta \, )^2\,
\biggl( \frac {\pi }{|\gamma |}\biggr) ^2 \bigl\| \phi ^{\, {\mathcal K}_{\beta }}
\bigr\| ^2\, +
$$ $$
v(K)\, \sum\limits_{N\, \in \, \Lambda ^*\backslash \, {\mathcal K}
_{\beta }}G_N^2(k;\varkappa )\, \| \phi _N\| ^2\biggr)
$$
holds.}
\vskip 0.2cm

Theorem 8 also follows from Theorem 7 in view of the uniform approximation of the
functions $\widehat V^{(s)}$ and $A$ by trigonometric polynomials with the period
lattice $\Lambda $.
\vskip 0.2cm

{\bf Corollary.} {\it Let $\widehat V^{(s)}:{\mathbb R}^n\to {\mathcal L}^{(s)}_M$, 
$s=0,1$, and $A:{\mathbb R}^n\to {\mathbb C}^n$ be continuous periodic functions 
with the period lattice $\Lambda \subset {\mathbb R}^n$, $n\geq 3$, let $A_0=0$, and 
let condition $\mathrm {(6)}$ with $\widetilde \theta \in [0,1)$ hold for some 
vector $\gamma \in \Lambda \backslash \{ 0\} $ $\mathrm ($$e=|\gamma |^{-1}\gamma 
$$\mathrm )$ and a measure $\mu \in {\mathcal M}_h\, $, $h>0$. Then there are 
numbers $c_9=c_9 \, (\gamma ,h,\mu ;\widehat V)>0$ and $\varkappa _0=\varkappa _0\, 
(\gamma ,h,\mu ;\widehat V)>0$ such that for all $k\in {\mathbb R}^n$ with 
$(k,\gamma )=\pi $, all $\varkappa \geq \varkappa _0\, $, and all vector functions 
$\phi \in \widetilde H^1(K;{\mathbb C}^M)$, the inequality
$$
\| \widehat D(k+i\varkappa e)\phi \| ^2 \, \geq \, 
c_9\, v(K) \sum\limits_{N\, \in \, \Lambda ^*}G_N^2(k;\varkappa )\, \| \phi _N\| ^2
$$
is fulfilled.}

\end{document}